\documentclass[a4paper,11pt]{article}
\usepackage{jcappub} 
\usepackage{lineno}

\usepackage{graphicx}
\usepackage{dcolumn}
\usepackage{bm}
\usepackage{amssymb}
\usepackage{latexsym}
\usepackage{booktabs}
\usepackage{amsmath}
\usepackage{multirow}
\usepackage{url}
\usepackage{hyperref}
\usepackage{footnote}
\usepackage{float}
\usepackage{threeparttable}
\usepackage{color}
\usepackage{array}
\usepackage{enumerate}
\usepackage{adjustbox}

\usepackage{makecell}
\usepackage{diagbox}
\usepackage{epstopdf}
\usepackage{epsfig}
\usepackage{longtable}
\usepackage{supertabular}
\usepackage{algorithm}
\usepackage{pifont}
\usepackage{algorithmic}
\usepackage{changepage}
\usepackage{setspace}

\newcommand{\blue}[1]{\textcolor{blue}{#1}}
\begin{document}

\title{Revisiting the phenomenologically emergent dark energy model: is non-zero equation of state of dark matter favored by DESI DR2?}

\author[a]{Tian-Nuo Li,}
\author[a]{Yi-Min Zhang,}
\author[b]{Yan-Hong Yao,}
\author[a]{Guo-Hong Du,}
\author[c]{Peng-Ju Wu,}
\author[a,\ast]{Jing-Fei Zhang}
\author[a,d,e,\ast]{and Xin Zhang\note[$\ast$]{Corresponding author.}}
\affiliation[a]{Key Laboratory of Cosmology and Astrophysics (Liaoning Province) \& College of Sciences, Northeastern University, Shenyang 110819, China}
\affiliation[b]{Institute of Fundamental Physics and Quantum Technology, Department of Physics, School of Physical Science and Technology, Ningbo University, Ningbo, Zhejiang 315211, China}
\affiliation[c]{School of Physics, Ningxia University, Yinchuan 750021, China}
\affiliation[d]{MOE Key Laboratory of Data Analytics and Optimization for Smart Industry, Northeastern University, Shenyang 110819, China}
\affiliation[e]{National Frontiers Science Center for Industrial Intelligence and Systems Optimization, Northeastern University, Shenyang 110819, China}

\emailAdd{litiannuo@stumail.neu.edu.cn, zhangyimin@stumail.neu.edu.cn, yaoyanhong@nbu.edu.cn, duguohong@stumail.neu.edu.cn, wupengju@nxu.edu.cn, jfzhang@mail.neu.edu.cn, zhangxin@mail.neu.edu.cn}

\abstract{The nature of dark matter remains one of the most fundamental and unresolved questions in modern cosmology. In most cosmological models, dark matter is typically modeled as pressureless dust with an equation of state (EoS) parameter $w_{\rm dm} = 0$. However, there is no fundamental theoretical reason to exclude the possibility of a non-zero dark matter EoS parameter. In this work, we explore the possibility of a non-zero dark matter EoS within the phenomenologically emergent dark energy (PEDE) model, given its simplicity and proven ability to alleviate the Hubble tension. We perform observational constraints by using the latest baryon acoustic oscillation data from DESI DR2, the cosmic microwave background (CMB) data from Planck, and the type Ia supernova data from DESY5 and PantheonPlus. From our analysis, we observe that a negative dark matter EoS parameter is preferred in all scenarios. Specifically, the CMB+DESI+DESY5 data yields $w_{\mathrm{dm}} = -0.00093 \pm 0.00032$, deviating from zero at approximately the $3\sigma$ level. However, this deviation is likely driven by unidentified systematics or inconsistencies in the DESY5 data, with the deviation decreasing to $2\sigma$ when using PantheonPlus data. Meanwhile, a negative $w_{\rm dm}$ would increase the Hubble tension due to the positive degeneracy between $w_{\rm dm}$ and $H_0$. Furthermore, Bayesian evidence suggests that the $\Lambda$CDM model is strongly preferred over the PEDE+$w_{\rm dm}$ model. These analyses illustrate that it is not possible to both support a non-zero dark matter component within the PEDE model and alleviate the Hubble tension simultaneously.}

\maketitle

\section{Introduction}


According to cosmological observational evidence and under the assumption that general relativity is the correct gravity theory, dark energy and dark matter are the two main components of the universe, comprising nearly 95\% of its total energy budget. Although the nature of dark energy and dark matter remains uncertain, we currently have simple and popular candidates for both, namely the cosmological constant ($\Lambda$) and cold dark matter (CDM). It is in this vein that the $\Lambda$CDM cosmological model, long regarded as the standard paradigm of modern cosmology, has demonstrated remarkable success in accounting for a wide array of cosmological observations, such as the cosmic microwave background (CMB), baryon acoustic oscillation (BAO), and type Ia supernova (SN). 

However, as the precision of cosmological parameters improves, discrepancies have emerged between the early-universe measurements assuming the $\Lambda$CDM model and those from late-universe direct measurements, particularly concerning the Hubble constant $H_0$. Specifically, the Planck collaboration inferred $H_0=67.36\pm0.54$ ${\rm km}~{\rm s}^{-1}~{\rm Mpc}^{-1}$ from the CMB data \cite{Planck:2018vyg}, but the SH0ES collaboration yielded $H_0 = 73.04 \pm 1.04$ ${\rm km}~{\rm s}^{-1}~{\rm Mpc}^{-1}$ using local distance ladder measurements based on Cepheid-calibrated supernovae \cite{Riess:2021jrx}. The discrepancy between the two $H_0$ values has a statistical significance exceeding $5\sigma$, which is referred to as the Hubble tension problem. In recent years, the Hubble tension has been extensively discussed \cite{Bernal:2016gxb,DiValentino:2017iww,Guo:2018ans,Vagnozzi:2019ezj,DiValentino:2019ffd,Cai:2021wgv,Vagnozzi:2021gjh,Vagnozzi:2021tjv,Escudero:2022rbq,Zhao:2022yiv,James:2022dcx,Jin:2022qnj,Song:2022siz,Vagnozzi:2023nrq,Zhang:2023gye,Pierra:2023deu,Jin:2023sfc,Huang:2024gfw,Huang:2024erq,Xiao:2024nmi,Han:2025fii,Song:2025ddm}; see also refs.~\cite{DiValentino:2021izs,Kamionkowski:2022pkx} for reviews. Additionally, the $\Lambda$CDM model suffers from two major problems related to the cosmological constant, namely the ``fine-tuning" and ``cosmic coincidence" \cite{Sahni:1999gb,Bean:2005ru}. These issues collectively indicate the need to explore other physical mechanisms beyond $\Lambda$CDM, such as modified gravity models \cite{Sotiriou:2008rp,Cai:2015emx,Nojiri:2017ncd}, dynamical dark energy models \cite{Boisseau:2000pr,Chevallier:2000qy,Kamenshchik:2001cp,Linder:2002et}, holographic dark energy models \cite{Huang:2004wt,Wang:2004nqa,Zhang:2005hs,Zhang:2005yz,Zhang:2007sh,Zhang:2009un,Li:2012spm,Landim:2015hqa,Zhang:2015rha,Wang:2016och,Drepanou:2021jiv,Wang:2023gov,Han:2024sxm}, and interacting dark energy models \cite{Farrar:2003uw,Zhang:2005rg,Zhang:2005rj,Zhang:2004gc,Cai_2005,Wang:2006qw,Zhang:2012uu,Costa:2013sva,Li:2014cee,Li:2014eha,Li:2015vla,Feng:2016djj,Pan:2019gop,Pan:2020zza,DiValentino:2020kpf,Yao:2023jau,Li:2023gtu,Han:2023exn,Giare:2024ytc,CosmoVerseNetwork:2025alb}.

{Specifically, motivated by the current status of cosmological observations and the significant tensions in the estimated values of some key parameters within the $\Lambda$CDM model, Li and Shafieloo \cite{Li:2019yem} proposed a novel dark energy model called the phenomenologically emergent dark energy (PEDE). Keeping in mind the idea of a minimal cosmology, the PEDE model shares the same parameters as the $\Lambda$CDM model. However, the cause of cosmic acceleration is a function that exhibits a hyperbolic tangent behavior with an embedded logarithmic function. An interesting feature of the PEDE function is that it is negligible at early times, meaning it does not impact the physical processes of the early universe, but its contribution gradually emerges at later times, producing the observed late-time acceleration. In ref.~\cite{Li:2019yem}, the results of an analysis based on CMB, BAO, and SN datasets show that the PEDE model significantly outperforms $\Lambda$CDM, demonstrating its potential to alleviate the Hubble tension. Subsequently, Pan \emph{et al.} \cite{Pan:2019hac} primarily used CMB observational data to demonstrate the alleviation of the Hubble tension at the $1\sigma$ confidence level.} Due to its simplicity and ability to alleviate the Hubble tension, the PEDE framework has been extensively discussed and extended by researchers \cite{Yang:2020ope,Li:2020ybr,Hernandez-Almada:2020uyr,DiValentino:2021rjj,Alestas:2021luu,Yang:2021eud,Yang:2022kho,Liu:2022mpj,Keeley:2022ojz,John:2023fsy,Cozzumbo:2024vxw,Hernandez-Almada:2024ost,Wang:2024dcx,Liu:2024dlf,Liu:2025mub,Manoharan:2025uix}. Interestingly, Yao \emph{et al.} \cite{Yao:2023ybs} considered a free dark matter equation of state (EoS) parameter within the PEDE framework, referred to as the PEDE+$w_{\rm dm}$ model, and found a preference for a negative dark matter EoS parameter at the $2\sigma$ confidence level.

In fact, while the dark matter EoS parameter is commonly set to zero in most cosmological models (indicating that dark matter is pressureless or cold), this assumption faces both observational and theoretical challenges \cite{Bullock:2017xww}. Therefore, a more general approach is to treat the dark matter EoS parameter as a free parameter, and it is particularly important to use observational data to examine whether it is consistent with zero, as demonstrated by several relevant studies (see, e.g., refs. \cite{Muller:2004yb,Kumar:2012gr,Murgia:2017lwo,Gariazzo:2017pzb,Murgia:2018now,Kopp:2018zxp,Schneider:2018xba,Kumar:2019gfl,Ilic:2020onu,Najera:2020smt,Pan:2022qrr,Yao:2023qve,Liu:2025mob}).

Recently, the measurements of BAO from the second data release (DR2) of the Dark Energy Spectroscopic Instrument (DESI) have significantly enhanced the precision of cosmological constraints on the $\Lambda$CDM model, the $w$CDM model, and the $w_0w_a$CDM model \cite{DESI:2025zgx}. Specifically, the combination of DESI DR2 BAO data with CMB and SN data indicates a $4.2\sigma$ preference for dynamical dark energy within the $w_0w_a$CDM model, providing stronger evidence compared to the first data release (DR1) \cite{DESI:2025zgx,DESI:2024mwx}. Previously, DR1 had attracted a significant number of researchers seeking to constrain various aspects of cosmological physics, as seen in refs. \cite{Li:2024qso,Du:2024pai,Giare:2024gpk,Dinda:2024ktd,Jiang:2024viw,Jiang:2024xnu,Escamilla:2024ahl,Sabogal:2024yha,Li:2024qus,Wu:2024faw,Li:2024bwr,Odintsov:2024woi,Li:2025owk,Du:2025iow,Feng:2025mlo,Wu:2025wyk}, with the recent DR2 continuing this trend \cite{DESI:2025fii,Ormondroyd:2025iaf,Pan:2025psn,Pang:2025lvh,Wang:2025ljj,Kessler:2025kju,Cheng:2025lod,Li:2025cxn,Yang:2025mws,Pan:2025qwy,You:2025uon,Silva:2025hxw,Tyagi:2025zov,Santos:2025wiv,Ling:2025lmw,Yashiki:2025loj,Cai:2025mas,Wang:2025dtk,vanderWesthuizen:2025iam,Luciano:2025elo,Ye:2025ark,Li:2025htp,Du:2025xes,Li:2025muv}. In particular, this has stimulated research into scenarios involving non-zero dark matter EoS \cite{Yang:2025ume,Chen:2025wwn,Wang:2025zri,Kumar:2025etf,Abedin:2025dis,Khoury:2025txd,Araya:2025rqz,Li:2025dwz,Yao:2025kuz,Yao:2025twv}. For example, Kumar \emph{et al.} \cite{Kumar:2025etf} investigated the potential deviations from CDM and found that the current observational data favor a non-zero dark matter EoS parameter at approximately the $2\sigma$ confidence level.

In this context, it is necessary to investigate the impact of this new dataset on the PEDE model with a free dark matter EoS parameter. In this work, we use the latest DESI DR2 BAO data, CMB data from Planck, and SN data from DESY5 and PantheonPlus to constrain the PEDE and PEDE+$w_{\rm dm}$ models. Our motivation is to explore whether a non-zero dark matter EoS parameter is favored by the current observational data. Furthermore, we study the effect of the PEDE model with a free dark matter EoS parameter on alleviating the Hubble tension. Finally, we use Bayesian evidence to evaluate whether the PEDE and PEDE+$w_{\rm dm}$ models are favored by the current observational data relative to the $\Lambda$CDM model.

This work is organized as follows. In section~\ref{sec2}, we briefly introduce the PEDE+$w_{\rm dm}$ model, as well as the cosmological data used in this work. In section~\ref{sec3}, we report the constraint results and make some relevant discussions. The conclusion is given in section~\ref{sec4}.

\section{Methodology and data}\label{sec2}

\subsection{PEDE+$w_{\rm dm}$ model}\label{sec2.1}

We assume a spatially flat, homogeneous, and isotropic spacetime, which is described by the Friedmann-Robertson-Walker (FRW) metric. Moreover, we assume that the gravitational sector of the universe is described by general relativity, with matter minimally coupled to gravity. Additionally, we consider that there are no non-gravitational interactions between any of the fluids, and the universe consists of radiation, baryons, non-CDM, and PEDE. Consequently, the dimensionless Hubble parameter is given by
\begin{equation}
\frac{H^2(z)}{H_0^2} = \Omega_{\rm r0}(1+z)^4 + \Omega_{\rm dm0}(1+z)^{3(1+w_{\rm dm})} + \Omega_{\rm b0}(1+z)^3 + \Omega_{\rm de0} f(z),
\end{equation}
where $H(z)$ is the Hubble parameter, $\Omega_{\rm r0}$, $\Omega_{\rm dm0}$, $\Omega_{\rm b0}$, and $\Omega_{\rm de0}$ are the density parameters for radiation, non-CDM, baryons, and PEDE, respectively. 
The function $f(z)$ is defined as \cite{Li:2019yem}
\begin{equation}
f(z) = 1 - \tanh(\log_{10}(1+z)).
\end{equation}

Since there are no interactions between the fluids, we can obtain the EoS for PEDE as
\begin{equation}
  w_{\rm de}(z) = -1 - \frac{1}{3 \ln 10} \left[ 1 + \tanh(\log_{10}(1+z)) \right].
\end{equation}
It is evident from this equation that the EoS for PEDE exhibits symmetry. Specifically, in the distant past (i.e., $z \rightarrow \infty$), we have $w_{\rm de} \rightarrow -1 - 2/(3 \ln 10)$, while in the distant future (i.e., $z \rightarrow -1$), $w_{\rm de} \rightarrow -1$. At the present time (i.e., $z = 0$), $w_{\rm de} \rightarrow -1 - 1/(3 \ln 10)$, indicating the behavior of phantom dark energy.

In the conformal Newtonian gauge, the perturbed FRW metric takes the form
\begin{equation}
  ds^2 = a^2(\tau) \left[ -(1+2\psi) d\tau^2 + (1-2\phi) d\vec{r}^2 \right],
\end{equation}
where $\psi$ and $\phi$ represent the metric potentials, and $\vec{r}$ is the spatial coordinate. From the first-order perturbations of the conserved stress-energy-momentum tensor, we derive the continuity and Euler equations for dark matter and PEDE in Fourier space

\begin{equation}
 \delta^{\prime}_{\rm dm}= -(1+w_{\rm dm}) \left(\theta_{\rm dm} - 3 \phi^{\prime} \right)
 -3 \mathcal{H} \delta_{\rm dm} (c^2_{\rm s,dm} - w_{\rm dm}) - 9 (1+w_{\rm dm})(c^2_{\rm s,dm} - c^2_{\rm a,dm})\mathcal{H}^2 \frac{\theta_{\rm dm}}{k^2},
\end{equation}
\begin{equation}
\theta^{\prime}_{\rm dm}=-(1-3 c^2_{\rm s,dm}) \mathcal{H} \theta_{\rm dm}  + \frac{c^2_{\rm s,dm}}{1+w_{\rm dm}}k^2 \delta_{\rm dm} + k^2\psi.
\end{equation}

\begin{equation}
 \delta^{\prime}_{\rm de}= -(1+w_{\rm de}) \left(\theta_{\rm de} - 3 \phi^{\prime} \right)
 -3 \mathcal{H} \delta_{\rm de} (c^2_{\rm s,de} - w_{\rm de}) - 9 (1+w_{\rm de})(c^2_{\rm s,de} - c^2_{\rm a,de})\mathcal{H}^2 \frac{\theta_{\rm de}}{k^2},
\end{equation}
\begin{equation}
\theta^{\prime}_{\rm de}=-(1-3 c^2_{\rm s,de}) \mathcal{H} \theta_{\rm de}  + \frac{c^2_{\rm s,de}}{1+w_{\rm de}}k^2 \delta_{\rm de} + k^2\psi.
\end{equation}

Here, the prime denotes the derivative with respect to conformal time, $\mathcal{H}$ is the conformal Hubble parameter, and $k$ is the magnitude of the wavevector. $\delta_{\rm de}$($\delta_{\rm dm}$) and $\theta_{\rm de}$($\theta_{\rm dm}$) represent the density and velocity divergence perturbations for dark energy (dark matter), respectively, and $c_{\rm s,de}$($c_{\rm s,dm}$) and $c_{\rm a,de}$($c_{\rm a,dm}$) are the sound speed and adiabatic sound speed for dark energy (dark matter). It is worth emphasizing that we fix $c^2_{\mathrm{s,de}} = 1$. For simplicity, and in view of the fact that dark matter governs the formation of large-scale structures in the universe, the sound speed of non-CDM is set to $c^2_{\mathrm{s,dm}} = 0$.

\subsection{Cosmological data}\label{sec2.2}

A summary of the cosmological parameters sampled and the priors applied across different models is presented in table~\ref{tab1}. For the $\Lambda$CDM and PEDE models analysis, we use six key cosmological parameters: the physical densities of baryons ($\omega_{\rm b}$) and dark matter ($\omega_{\rm dm}$), the angular size of the horizon at the last scattering surface ($\theta_{\rm s}$), the optical depth ($\tau$), the amplitude of primordial scalar perturbation ($\log(10^{10}A_{\rm s})$), and the scalar spectral index ($n_{\rm s}$). For the PEDE+$w_{\rm dm}$ model analysis, we include the usual six $\Lambda$CDM parameters along with an additional dimensionless parameter: $w_{\rm dm}$.

We compute the theoretical model using a modified version of the {\tt CLASS} code \cite{Lesgourgues:2011re,Blas:2011rf}. We perform Markov Chain Monte Carlo (MCMC) \cite{Lewis:2002ah,Lewis:2013hha} 
analyses using the publicly available sampler {\tt Cobaya}\footnote{\url{https://github.com/CobayaSampler/cobaya}.} \cite{Torrado:2020dgo} and assess the convergence of the MCMC chains using the Gelman-Rubin statistics quantity $R - 1 < 0.02$ \cite{Gelman:1992zz}. The MCMC chains are analyzed using the public package {\tt GetDist}\footnote{\url{https://github.com/cmbant/getdist}.} \cite{Lewis:2019xzd}. We use the current observational data to constrain these models and obtain the best-fit values and the $1$--$2\sigma$ confidence level ranges for the parameters of interest \{$H_{0}$, $\Omega_{\mathrm{m}}$, $w_{\rm dm}$\}. Next, we present a comprehensive overview of the observational data employed in this work.

\begin{table}[t]
\caption{Flat priors on the main cosmological parameters constrained in this paper.}
\begin{center}
\renewcommand{\arraystretch}{1.5}
\begin{tabular}{c@{\hspace{0.4cm}}@{\hspace{0.4cm}} c @{\hspace{0.4cm}} c }
\hline\hline
\textbf{Model}       & \textbf{Parameter}       & \textbf{Prior}\\
\hline
$\Lambda$CDM/PEDE           & $\omega_{\rm b}$                     & $\mathcal{U}$[0.005\,,\,0.1] \\
                    &$\omega_{\rm dm}$                     & $\mathcal{U}$[0.01\,,\,0.99] \\
                    &$100\theta_{\rm s}$                 & $\mathcal{U}$[0.5\,,\,10] \\
                    & $\tau$                                   & $\mathcal{U}$[0.01\,,\,0.8] \\
                    & $\log(10^{10}A_{\rm s})$                 & $\mathcal{U}$[1.61\,,\,3.91] \\
                    & $n_{\rm s}$                                  & $\mathcal{U}$[0.8\,,\, 1.2] \\
\hline
PEDE+$w_{\rm dm}$       & $w_{\rm dm}$                                & $\mathcal{U}$[-0.1\,,\, 0.1] \\

\hline\hline
\end{tabular}
\label{tab1}
\end{center}	
\end{table}

\begin{itemize}

\item Cosmic Microwave Background. The analysis involves measurements of the Planck CMB temperature anisotropy and polarization power spectra, their cross-spectra, as well as the Planck lensing power spectrum. This work employs the following four components of CMB likelihoods: (i) the power spectra of temperature and polarization anisotropies, $C_{\ell}^{TT}$, $C_{\ell}^{TE}$, and $C_{\ell}^{EE}$, at small scales ($\ell > 30$), are obtained from measurements using the Planck \texttt{CamSpec} likelihood~\cite{Planck:2018vyg,Efstathiou:2019mdh,Rosenberg:2022sdy}; (ii) the spectrum of temperature anisotropies, $C_{\ell}^{TT}$, at large scales ($2 \leq \ell \leq 30$), is obtained from measurements using the Planck \texttt{Commander} likelihood~\cite{Planck:2018vyg,Planck:2019nip}; (iii) the spectrum of E-mode polarization, $C_{\ell}^{EE}$, at large scales ($2 \leq \ell \leq 30$), is obtained from measurements using the Planck \texttt{SimAll} likelihood~\cite{Planck:2018vyg,Planck:2019nip}; (iv) the CMB lensing likelihood, with the latest and most precise data coming from the NPIPE PR4 Planck CMB lensing reconstruction\footnote{\url{https://github.com/carronj/planck_PR4_lensing}.} \cite{Carron:2022eyg}. We denote the combination of these likelihoods as \textbf{``CMB"}.

\item Baryon Acoustic Oscillation. The BAO measurements from DESI DR2 include tracers of the bright galaxy sample, luminous red galaxies, emission line galaxies, quasars, and the Lyman-$\alpha$ forest. These tracers are described through the transverse comoving distance $D_{\mathrm{M}}/r_{\mathrm{d}}$, the angle-averaged distance $D_{\mathrm{V}}/r_{\mathrm{d}}$, and the Hubble horizon $D_{\mathrm{H}}/r_{\mathrm{d}}$, where  $r_{\mathrm{d}}$ is the comoving sound horizon at the drag epoch. We use measurements detailed in Table IV of ref.~\cite{DESI:2025zgx}. We denote this full dataset as \textbf{``DESI"}.

\item Type Ia Supernova. We adopt SN data from two compilations: (i) the PantheonPlus comprises 1550 spectroscopically confirmed type Ia supernovae (SNe) from 18 different surveys, spanning the redshift range of $0.01 < z < 2.26$ \cite{Brout:2022vxf}\footnote{\url{https://github.com/PantheonPlusSH0ES/DataRelease}.}; (ii) the Dark Energy Survey collaboration released part of the full 5 yr data based on a new, homogeneously selected sample of 1635 photometrically classified SNe (with redshifts in the range $0.1 < z < 1.3$), complemented by 194 low-redshift SNe from the CfA3~\cite{Hicken:2009df}, CfA4~\cite{Hicken:2012zr}, CSP~\cite{Krisciunas:2017yoe}, and Foundation~\cite{Foley:2017zdq} samples (with redshifts in the range $0.025 < z < 0.1$), totaling 1829 SNe \cite{DES:2024jxu}.\footnote{\url{https://github.com/des-science/DES-SN5YR}.} We label the two sets of SN data as \textbf{``PantheonPlus''} and \textbf{``DESY5''}, respectively.

\end{itemize}

\begin{table*}[t]
\renewcommand\arraystretch{1.5}
\centering
\caption{Fitting results ($1\sigma$ confidence level) in the $\Lambda$CDM, PEDE, and PEDE+$w_{\rm dm}$ models from the CMB, CMB+DESI, CMB+DESI+DESY5, and CMB+DESI+PantheonPlus data. Here, $H_{0}$ is expressed in units of ${\rm km}~{\rm s}^{-1}~{\rm Mpc}^{-1}$.}
\label{tab2}
\resizebox{\textwidth}{!}{%
\begin{tabular}{lccccc} 
\hline
\hline
Model&Parameter & CMB& CMB+DESI & CMB+DESI+DESY5 & CMB+DESI+PantheonPlus \\ 
\hline
$\Lambda$CDM	& $H_{0}$ & $67.23\pm 0.48$ & $68.19\pm0.27$ & $68.00\pm0.27$ & $68.10\pm 0.28$ \\
& $\Omega_{\mathrm{m}}$ & $0.3155\pm0.0066$ & $0.3024\pm0.0035$  & $0.3049\pm0.0035$& $0.3036\pm 0.0035$\\

PEDE & $H_{0}$ & $72.44\pm0.62$ & $72.08\pm0.33$ & $71.34\pm0.31$ & $71.53\pm 0.33$\\
& $\Omega_{\mathrm{m}}$ & $0.2713\pm 0.0065$ & $ 0.2750\pm 0.0035$ & $0.2830\pm0.0034$& $0.2809\pm 0.0035$\\

PEDE+$w_{\rm dm}$ & $H_{0}$ & $71.50^{+3.10}_{-3.90}$ & $ 71.83\pm 0.46 $ & $ 70.62\pm0.44$ & $ 70.93\pm 0.45$ \\
& $\Omega_{\mathrm{m}}$ & $0.2840\pm0.0370$ & $0.2772\pm0.0045$ & $0.2897\pm0.0043$  & $0.2864\pm 0.0043$ \\

& $w_{\rm dm}$ & $-0.00040\pm0.00150$ & $-0.00036\pm0.00034$ & $-0.00093\pm0.00032$ & $-0.00065\pm 0.00032$  \\

\hline
\end{tabular}
}
\end{table*}

\section{Results and discussions}\label{sec3}

\begin{figure*}[htbp]
\centering
\includegraphics[width=0.46\textwidth, height=0.305\textheight]{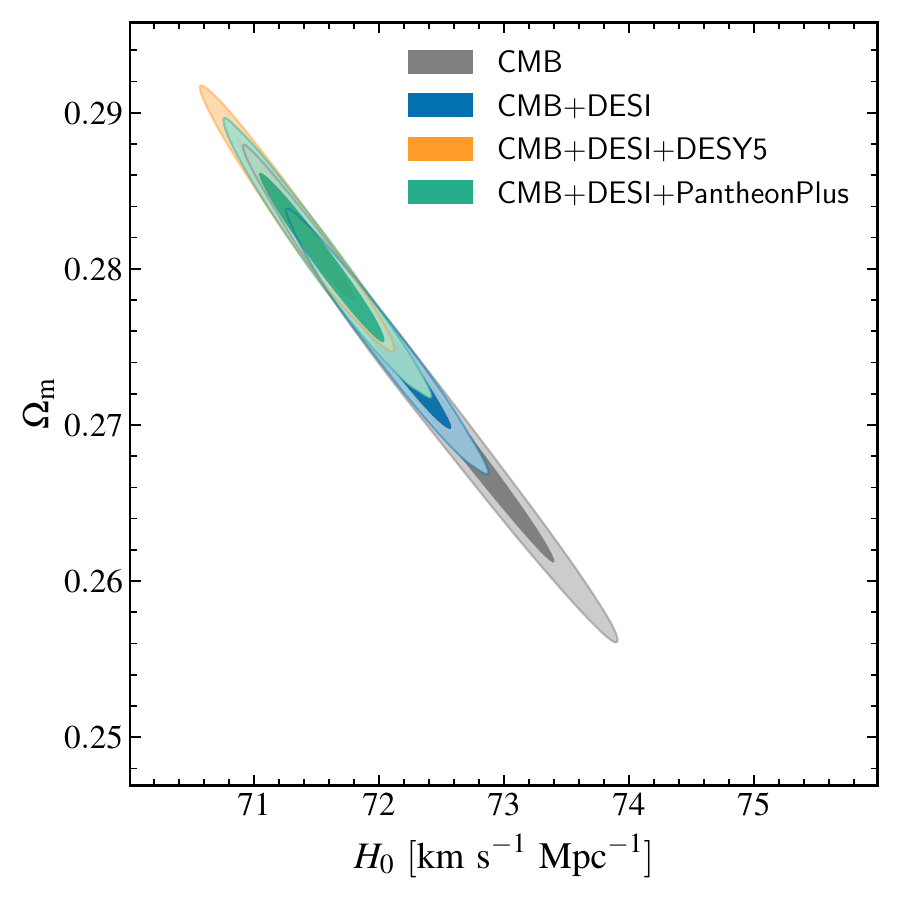} \hspace{1cm}
\includegraphics[width=0.44\textwidth, height=0.3\textheight]{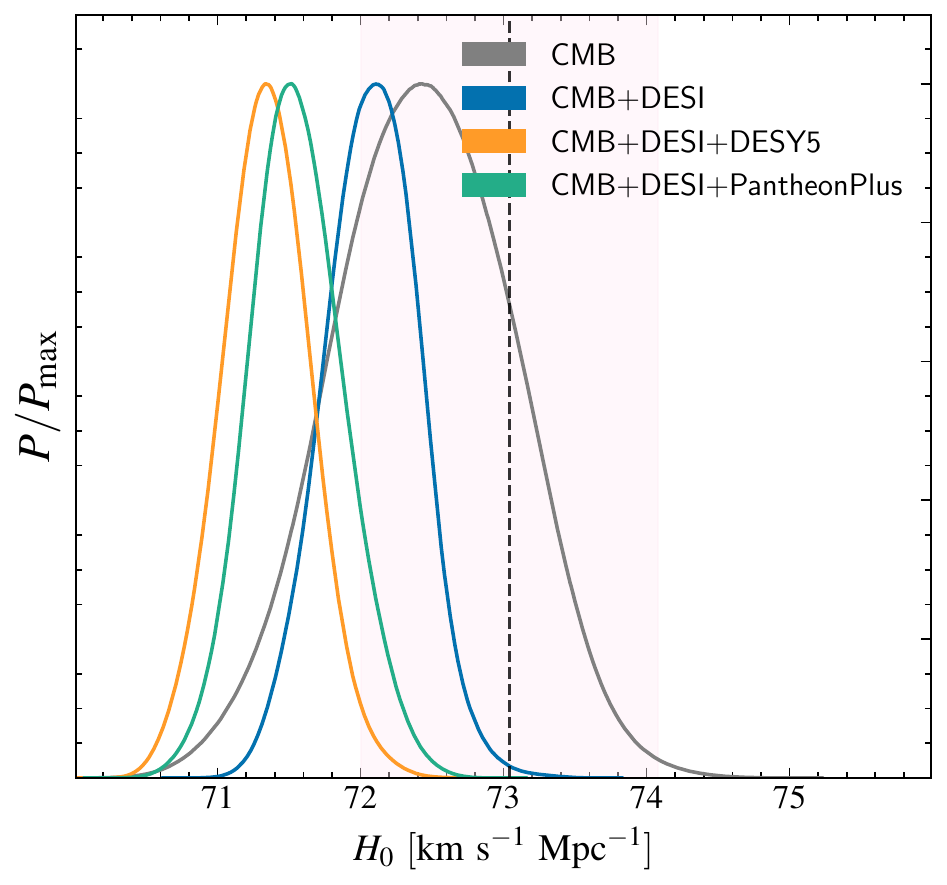}
\caption{\label{fig1} Constraints on the cosmological parameters using the CMB, DESI, DESY5, and PantheonPlus data in the PEDE model. \emph{Left panel}: Two-dimensional marginalized contours ($1\sigma$ and $2\sigma$ confidence levels) in the $H_0$--$\Omega_{\rm m}$ plane by using the CMB, CMB+DESI, CMB+DESI+DESY5, and CMB+DESI+PantheonPlus data in the PEDE model.  \emph{Right panel}: The marginalized one-dimensional posteriors on $H_0$ in PEDE model, from CMB, CMB+DESI, CMB+DESI+DESY5, and CMB+DESI+PantheonPlus as labelled. The light pink vertical band corresponds to the value of $H_0$ from the SH0ES team \cite{Riess:2021jrx}.}
\end{figure*}

\begin{figure*}[htbp]
\includegraphics[scale=0.7]{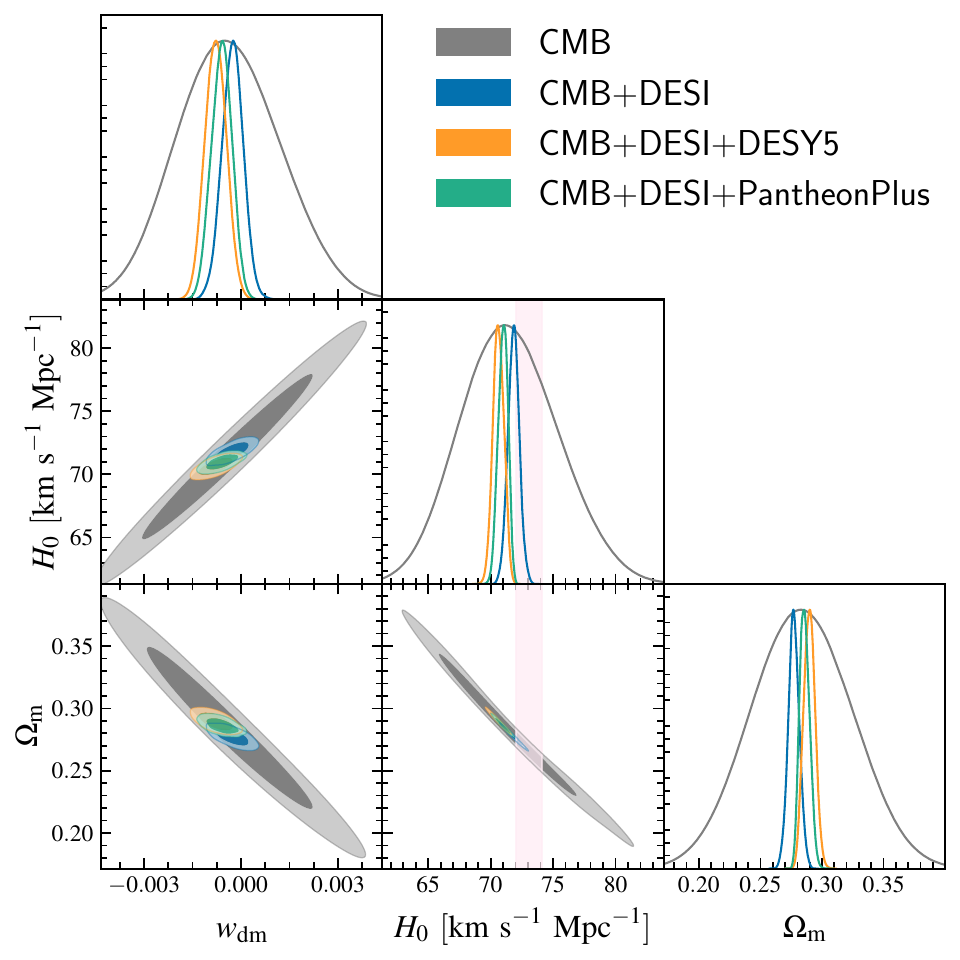}
\centering
\caption{\label{fig2} Constraints on the cosmological parameters using the CMB, CMB+DESI, CMB+DESI+DESY5, and CMB+DESI+PantheonPlus data in the PEDE+$w_{\rm dm}$ model. The light pink vertical band corresponds to the value of $H_0$ from the SH0ES team \cite{Riess:2021jrx}.}
\end{figure*}

\begin{figure}
\includegraphics[scale=0.6]{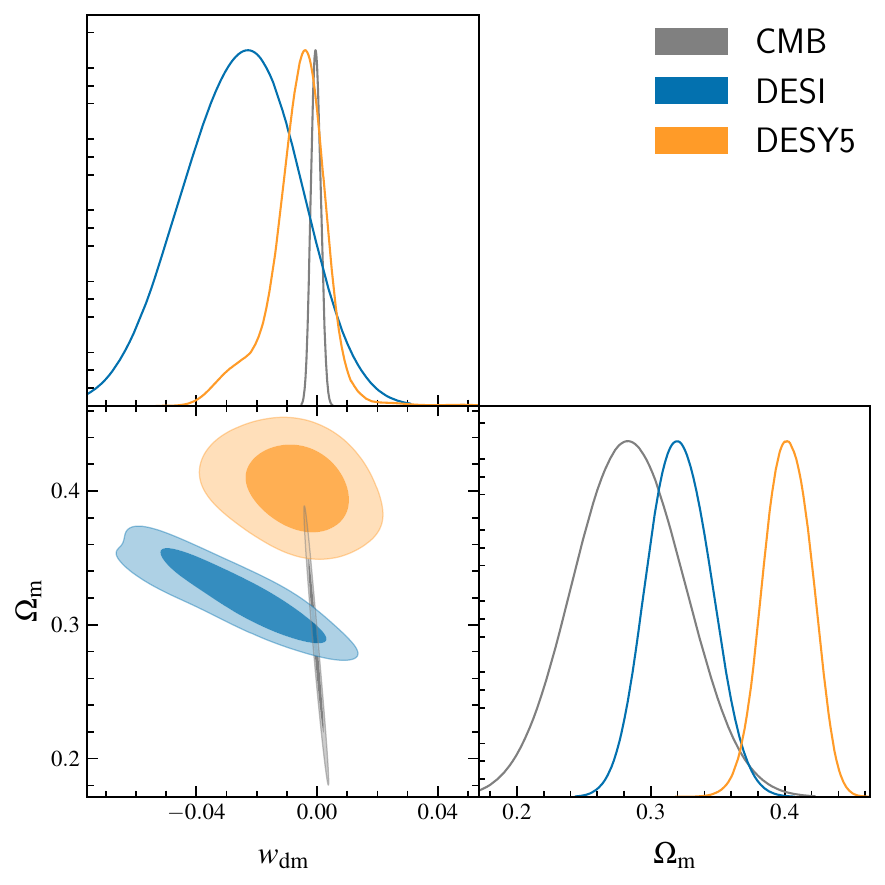}
\centering
\caption{\label{fig3} Constraints on the cosmological parameters using the CMB, DESI, and DESY5 data in the PEDE+$w_{\rm dm}$ model.}
\end{figure}

\begin{figure}
\includegraphics[scale=0.4]{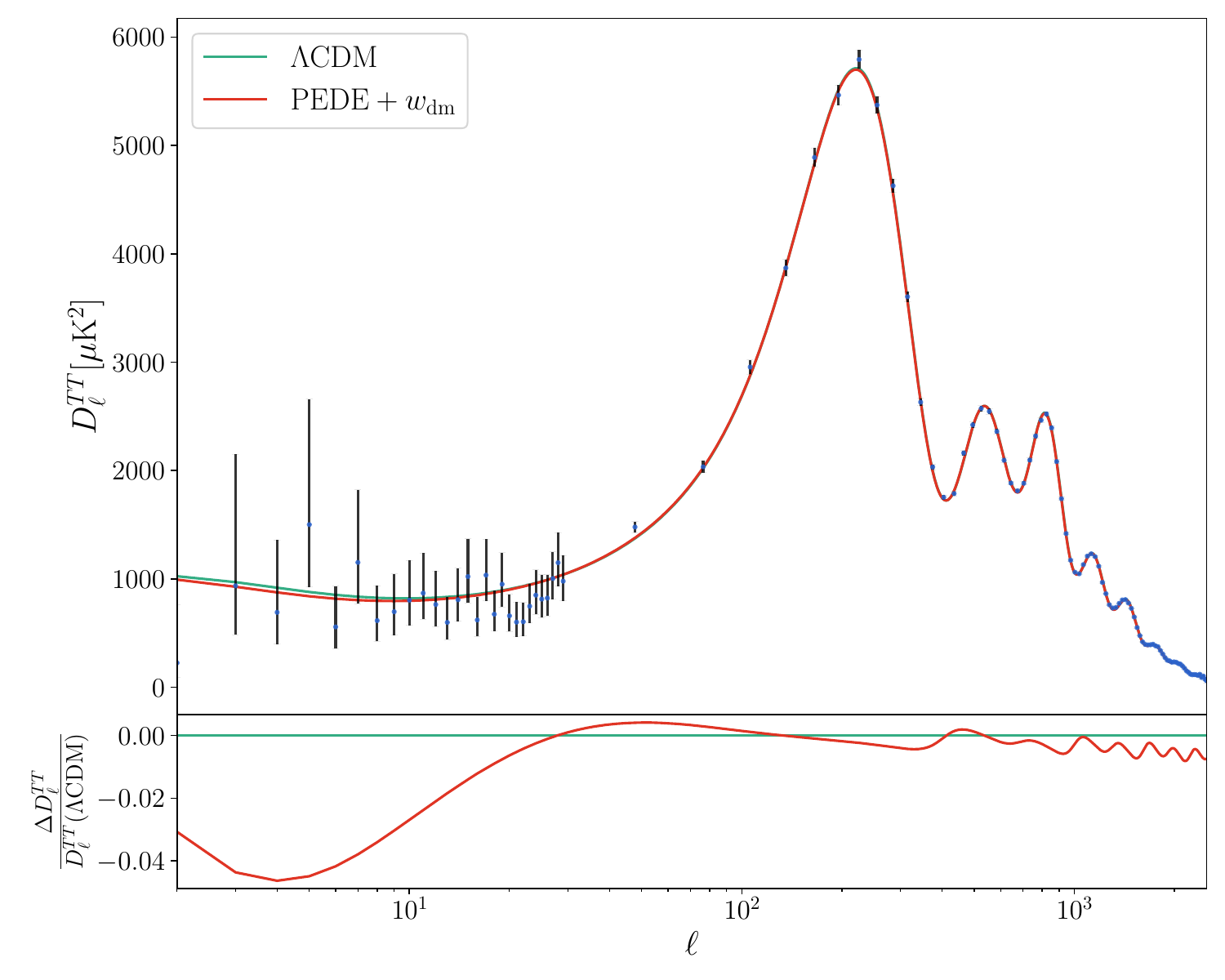}
\centering
\caption{\label{fig4} Comparison of the temperature power spectrum for the $\Lambda$CDM (green) and PEDE+$w_{\rm dm}$ (red) models, using the best-fit values inferred from the CMB alone analysis, over the Planck temperature power spectrum data points (black vertical bars with blue dots). We also present the fractional difference between the $\Lambda$CDM and PEDE+$w_{\rm dm}$ models.}
\end{figure}

\begin{figure}
\includegraphics[scale=0.45]{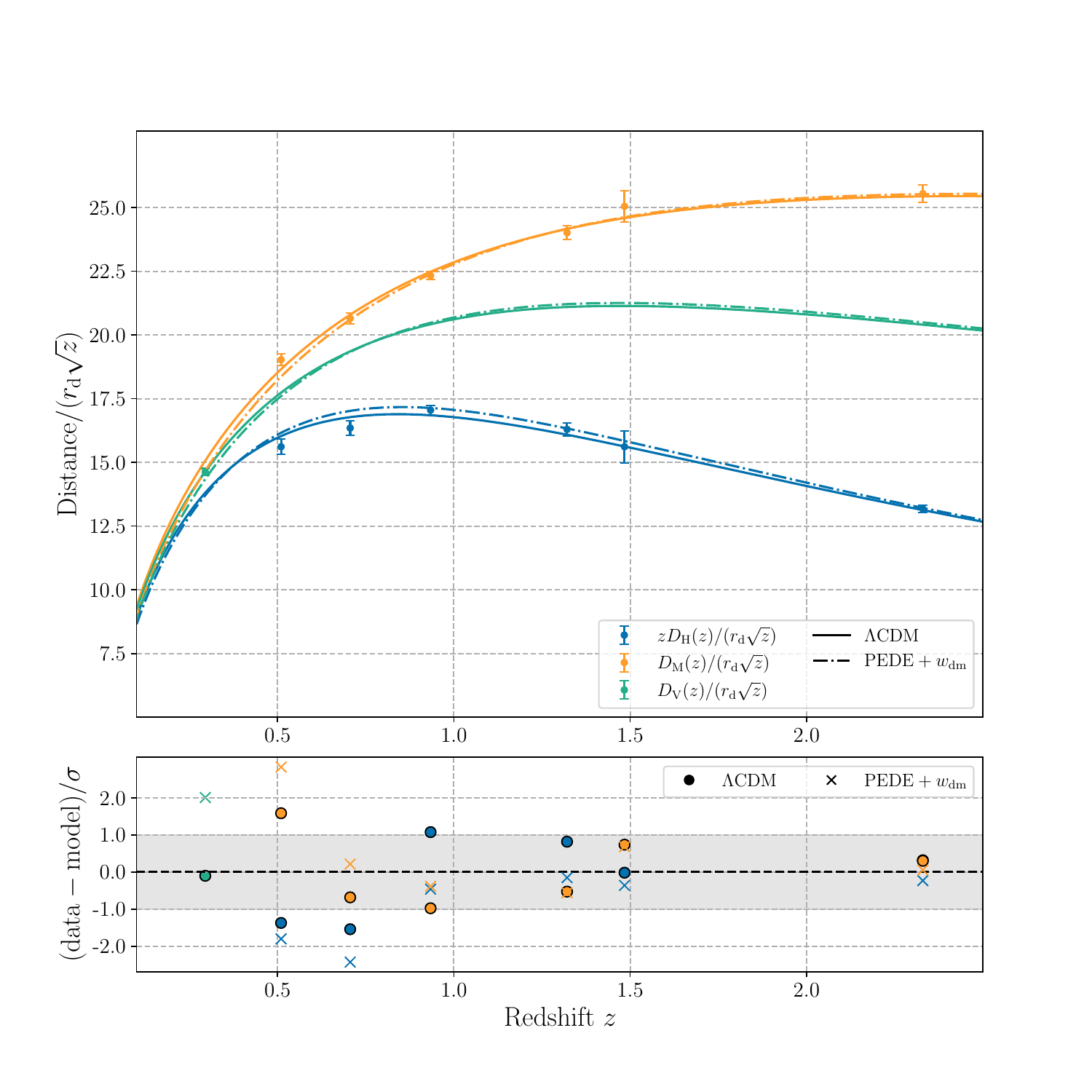}
\centering
\caption{\label{fig5} Best-fit predictions for distance-redshift relations for the $\Lambda$CDM and PEDE+$w_{\rm dm}$ models using CMB and DESI data. \emph{Upper panel}: Best-fit predictions for (rescaled) distance-redshift relations for $\Lambda$CDM (solid line) and  PEDE+$w_{\rm dm}$ (dotted-dashed line) obtained from the analysis of CMB+DESI data. The predictions encompass the three distinct types of distances probed by DESI BAO measurements, including $D_{\mathrm{V}}$ (green), $D_{\mathrm{M}}$ (orange), and $D_{\mathrm{H}}$ (light blue). The error bars represent $1\sigma$ uncertainties. \emph{Lower panel}: Difference between the model prediction and data-point for each BAO measurement, normalized by the observational uncertainties. The predictions for $\Lambda$CDM and PEDE+$w_{\rm dm}$ are shown by filled and \blue{cross-shaped} markers, respectively. The shaded region highlights the $±1\sigma$ range for reference.}
\end{figure}

\begin{figure*}[!htp]
\includegraphics[scale=0.45]{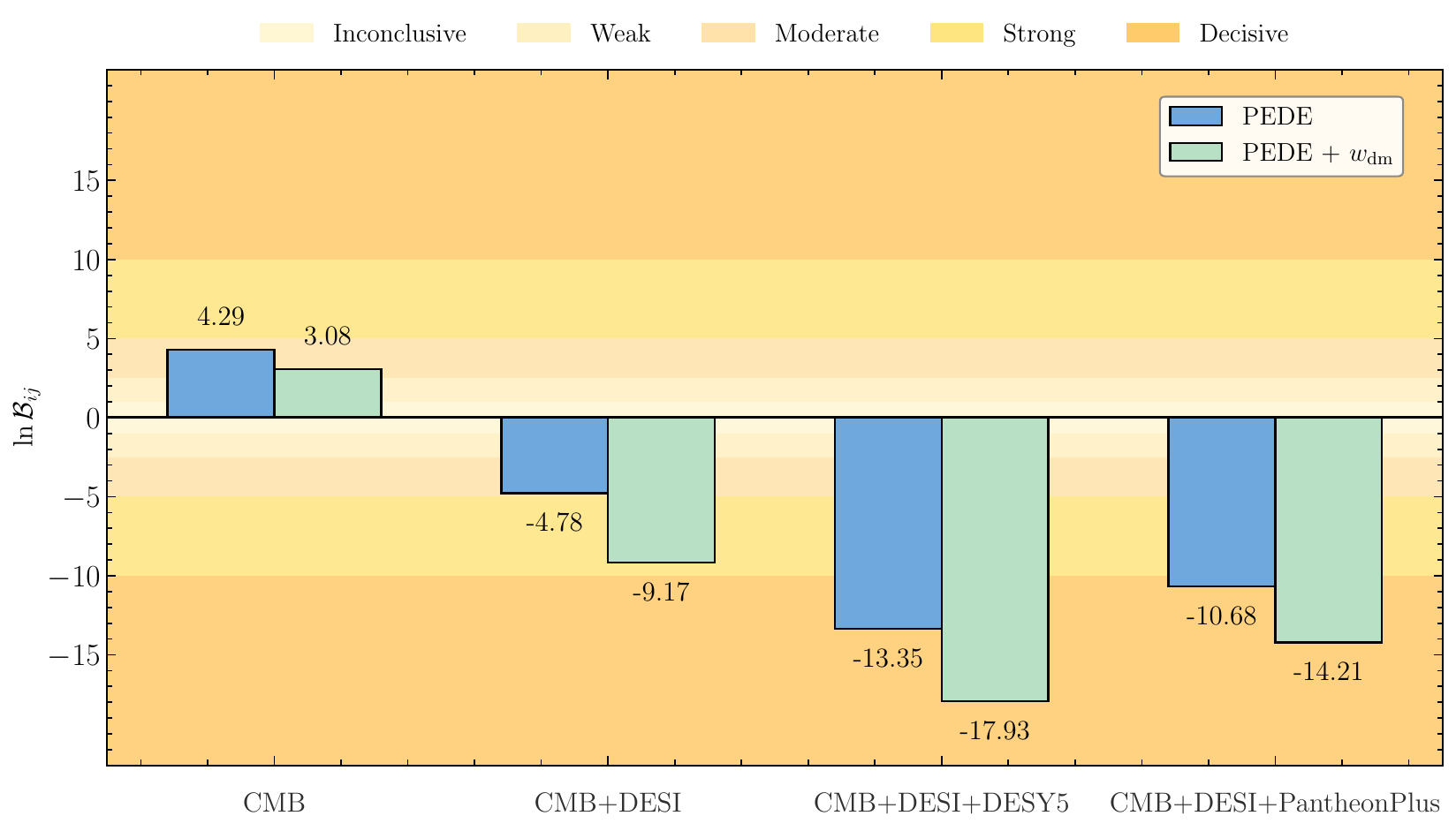}
\centering
\caption{\label{fig6} Comparison of the Bayesian evidence for the PEDE and PEDE+$w_{\rm dm}$ models relative to the $\Lambda$CDM model using current observational data. The Bayes factor $\ln \mathcal{B}_{ij}$ (where $i$ = PEDE or PEDE+$w_{\rm dm}$, $j$ = $\Lambda$CDM) and its strength according to Jeffreys' scale are used to assess the preference between models, where a negative value indicates a preference for the $\Lambda$CDM model.}
\end{figure*}

In this section, we shall report the constraint results of the cosmological parameters. We consider the PEDE and PEDE+$w_{\rm dm}$ models to perform a cosmological analysis using current observational data, including the DESI, CMB, DESY5, and PantheonPlus data. We show the $1\sigma$ and $2\sigma$ posterior distribution contours for various cosmological parameters in the PEDE and PEDE+$w_{\rm dm}$ models in figures~\ref{fig1}--\ref{fig3}. The $1\sigma$ errors for the marginalized parameter constraints are summarized in table~\ref{tab2}. We plot the best-fit curves for the CMB temperature power spectra of the $\Lambda$CDM and PEDE+$w_{\rm dm}$ models, obtained from the CMB data, as shown in figure~\ref{fig4}. We compare the best-fit predictions for the three different types of (rescaled) distances obtained by DESI BAO measurements using CMB+DESI data in the $\Lambda$CDM and PEDE+$w_{\rm dm}$ models, as shown in figure~\ref{fig5}. Finally, we compare $\ln \mathcal{B}_{ij}$ of the PEDE and PEDE+$w_{\rm dm}$ models relative to the $\Lambda$CDM model using the current observational data, as shown in figure~\ref{fig6}.

In the left panel of figure~\ref{fig1}, we present the constraint results of the PEDE model in the $H_0$--$\Omega_{\mathrm{m}}$ plane from the current observational data. For the CMB, CMB+DESI, CMB+DESI+DESY5, and CMB+DESI+PantheonPlus data, the constraint values of $H_0$ are $72.44 \pm 0.62~\rm km~s^{-1}~Mpc^{-1}$, $72.08 \pm 0.33~\rm km~s^{-1}~Mpc^{-1}$, $71.34 \pm 0.31~\rm km~s^{-1}~Mpc^{-1}$, and $71.53\pm 0.33~\rm km~s^{-1}~Mpc^{-1}$, respectively, and the corresponding values of $\Omega_{\mathrm{m}}$ are $0.2713 \pm 0.0065$, $0.2750 \pm 0.0035$, $0.2830 \pm 0.0034$, and $0.2809\pm 0.0035$, respectively. For all datasets, we obtain a higher $H_0$ value and a slightly lower $\Omega_{\mathrm{m}}$ value, relative to the $\Lambda$CDM model. This is because $\Omega_{\mathrm{m}}$ and $H_0$ also exhibit an anti-correlation within the PEDE model. In the right panel of figure~\ref{fig1}, we present the constraint results of the CMB, CMB+DESI, CMB+DESI+DESY5, and CMB+DESI+PantheonPlus data in the one-dimensional marginalized posterior distributions on $H_0$. When using CMB alone, the central value of $H_0$ for the PEDE model is in very close agreement with the $H_0$ estimate from SH0ES. When the DESI, DESY5, and PantheonPlus data are added, although the $H_0$ value slightly decreases, it remains higher than that obtained from the $\Lambda$CDM model. In the PEDE model, the values of $H_0$ obtained using the CMB, CMB+DESI, CMB+DESI+DESY5, and CMB+DESI+PantheonPlus data are in $0.49\sigma$, $0.88\sigma$, $1.57\sigma$, and $1.38\sigma$ tensions with SH0ES, respectively.

In figure~\ref{fig2}, we show the triangular plot of the constraint results for the PEDE+$w_{\rm dm}$ model using the CMB, CMB+DESI, CMB+DESI+DESY5, and CMB+DESI+PantheonPlus data. We find that $w_{\mathrm{dm}}$ and $H_0$ exhibit a strong positive correlation. When using CMB alone, the constraint results are $w_{\mathrm{dm}} = -0.00040 \pm 0.00150$ and $H_0 = 71.50^{+3.10}_{-3.90}~\rm km~s^{-1}~Mpc^{-1}$, with relatively weak constraints on $H_0$. Notably, no evidence for a deviation of $w_{\rm dm}$ from zero is found. We find that the Hubble constant in the PEDE+$w_{\rm dm}$ model decreases to $H_0 = 71.50^{+3.1}_{-3.9}~\rm km~s^{-1}~Mpc^{-1}$, compared to $H_0 = 72.44\pm 0.62~\rm km~s^{-1}~Mpc^{-1}$ in the PEDE model, due to the positive correlation between $w_{\rm dm}$ and $H_0$. 

When combining DESI with CMB, the constraint on $w_{\mathrm{dm}}$ is $-0.00036 \pm 0.00034$, indicating a preference for a negative dark matter EoS parameter, with the deviation of $w_{\mathrm{dm}}$ from zero reaching $1.1\sigma$ level. We find that the Hubble constant in the PEDE+$w_{\rm dm}$ model decreases to $H_0 = 71.83\pm 0.46~\rm km~s^{-1}~Mpc^{-1}$, compared to the PEDE model. As a result, the Hubble tension increases from $0.88\sigma$ in the PEDE model to $1.06\sigma$ in the PEDE+$w_{\rm dm}$ model. Furthermore, the combination of low-redshift data and CMB data can effectively break the cosmological parameter degeneracies. For example, CMB+DESI results in $\sigma(w_{\mathrm{dm}})=0.00034$, $\sigma(H_0)=0.46~\rm km~s^{-1}~Mpc^{-1}$, and $\sigma(\Omega_{\rm m})=0.0045$, which are 77.3\%, 86.9\%, and 87.8\% better than those of CMB, respectively. 

When combining the DESY5 and DESI data with CMB, the constraint on $w_{\mathrm{dm}}$ is $-0.00093 \pm 0.00032$, showing a larger deviation of $w_{\rm dm}$ from zero at approximately the $3\sigma$ level, further supporting a negative dark matter EoS parameter. Furthermore, the value of $H_0$ is $70.62 \pm 0.44~\rm km~s^{-1}~Mpc^{-1}$, and the Hubble tension has increased from $1.06\sigma$ to $2.14\sigma$ relative to the CMB+DESI combination. Meanwhile, compared to the value of $H_0$ in the PEDE model, the Hubble tension with SH0ES increases from $1.57\sigma$ to $2.14\sigma$. Thus, a more negative value of $w_{\rm dm}$ would exacerbate the Hubble tension, as suggested across all data combinations. The results obtained here indicate that we cannot simultaneously obtain evidence for non-CDM and alleviate the Hubble tension within the framework of this study.

It is worth noting that the $w_{\rm dm}$ constraint errors from CMB+DESI+DESY5 are essentially consistent with those from CMB+DESI, with the only difference being a shift in the central value, which leads to the deviation of $w_{\rm dm}$ from zero increasing from $1.1\sigma$ to approximately $3\sigma$. However, when using the CMB+DESI+PantheonPlus data, the constraint on $w_{\mathrm{dm}}$ is $-0.00065 \pm 0.00032$, with the deviation from zero decreasing from approximately $3\sigma$ to $2\sigma$ compared to the CMB+DESI+DESY5 data. These results suggest that the $3\sigma$ deviation observed with CMB+DESI+DESY5 is likely driven by DESY5 (unidentified supernova systematics \cite{DES:2025tir,Huang:2025som}) or by inconsistencies between the datasets. Therefore, we perform constraints on the PEDE+$w_{\rm dm}$ model using CMB, DESI, and DESY5 data separately, as shown in figure~\ref{fig3}. For the CMB, DESI, and DESY5 data, the constraint values of $w_{\rm dm}$ are $-0.0004 \pm 0.0015$, $-0.0250 \pm 0.0170$, and $0.0064^{+0.0110}_{-0.0065}$, respectively, while the corresponding values of $\Omega_{\mathrm{m}}$ are $0.284 \pm 0.037$, $0.321 \pm 0.021$, and $0.402 \pm 0.019$, respectively. We find that there are some discrepancies in the central values of $\Omega_{\mathrm{m}}$, and that $\Omega_{\mathrm{m}}$ and $w_{\rm dm}$ exhibit an anti-correlation. The discrepancy in $\Omega_{\mathrm{m}}$ between DESY5 and CMB is approximately $2.89\sigma$, which may influence the shift in $w_{\rm dm}$. {It should be emphasized that DESY5 alone cannot place effective constraints on parameters in the PEDE+$w_\mathrm{dm}$ model other than $w_\mathrm{dm}$ and $\Omega_{\rm m}$. Therefore, the relatively high $\Omega_{\rm m}$ reported by DESY5 is unlikely to be driven by projection effects of other parameters, while it is more likely to originate from unknown systematic errors in this data. See refs.~\cite{Notari:2024zmi,Efstathiou:2024xcq,Huang:2024qno,Colgain:2024ksa,Colgain:2024mtg,Tang:2024lmo} for detailed discussions on the unknown systematic errors of DESY5 and $\Omega_{\mathrm{m}}$ inconsistencies.}

In figure~\ref{fig4}, we present the best-fit curves for the CMB temperature power spectra of the $\Lambda$CDM model (green) and the PEDE+$w_{\rm dm}$ model (red), obtained from CMB data, along with the Planck temperature power spectrum data points. We also present the fractional difference between the $\Lambda$CDM and PEDE+$w_{\rm dm}$ models. We can clearly see that the difference primarily arises at lower multipoles, while the two models are nearly identical at higher multipoles. We observe that the amplitude at lower multipoles is slightly suppressed in PEDE+$w_{\rm dm}$ model, resulting in a better agreement with the CMB data.

In order to better understand the role played by DESI data in PEDE+$w_{\rm dm}$ model, we compare the theoretical distance predictions of $\Lambda$CDM and PEDE+$w_{\rm dm}$ models against the observed cosmic distances in figure~\ref{fig5}. Due to the relatively weak constraints provided by DESI alone, we combine DESI with CMB for our study. We analyze and compare the best-fit predictions from CMB+DESI for three different types of (rescaled) distances, including $D_{\mathrm{V}}$, $D_{\mathrm{M}}$, and $D_{\mathrm{H}}$ derived from DESI BAO measurements. In the upper panel of figure~\ref{fig5}, we observe that the theoretical distance predictions of the $\Lambda$CDM and PEDE+$w_{\rm dm}$ models exhibit certain deviations. In the lower panel of figure~\ref{fig5}, we observe that the quantities $D_{\mathrm{M}}(z)/(r_{\mathrm{d}}\sqrt{z})$ and $zD_{\mathrm{H}}(z)/(r_{\mathrm{d}}\sqrt{z})$ at $z = 0.51$ and $z = 0.71$ deviate from the theoretical distance predictions of the $\Lambda$CDM and PEDE+$w_{\rm dm}$ models. However, PEDE+$w_{\rm dm}$ model demonstrates greater success than the $\Lambda$CDM model in explaining $D_{\mathrm{M}}(z)/(r_{\mathrm{d}}\sqrt{z})$ at $z = 0.71$. Interestingly, we find that at $z > 1.1$, the results exhibit general agreement with the $\Lambda$CDM model, except for $zD_{\mathrm{H}}(z)/(r_{\mathrm{d}}\sqrt{z})$ at $z = 1.32$.

Finally, we adopt the Bayesian evidence selection criterion as a method for selecting the best model. Here, we use the publicly available code
{\tt MCEvidence}\footnote{\url{https://github.com/yabebalFantaye/MCEvidence}.} \cite{Heavens:2017hkr,Heavens:2017afc} to compute the Bayesian evidence of the models under consideration. The Bayesian evidence $Z$ is given by
\begin{equation}
Z = \int_{\Omega} P(D|\bm{\theta},M)P(\bm{\theta}|M)P(M)\ {\rm d}\bm{\theta},
\label{eq: lnZ}
\end{equation}
where $P(D|\bm{\theta},M)$ is the likelihood of the data $D$ given the parameters $\bm{\theta}$ and the model $M$, $P(\bm{\theta}|M)$ is the prior probability of $\bm{\theta}$ given $M$, and $P(M)$ is the prior of $M$. Then we calculate the Bayes factor $\ln \mathcal{B}_{ij} = \ln Z_i - \ln Z_j$ in logarithmic space, where $Z_i$ and $Z_j$ are the Bayesian evidence of two models.

Generally, the strength of model preference is commonly assessed using the Jeffreys scale \cite{Kass:1995loi,Trotta:2008qt}, which interprets the magnitude of the Bayes factor $\ln \mathcal{B}_{ij}$ as follows: if $\left|\ln \mathcal{B}_{ij}\right|<1$, the evidence is inconclusive; $1\le\left|\ln \mathcal{B}_{ij}\right|<2.5$ represents weak evidence; $2.5\le\left|\ln \mathcal{B}_{ij}\right|<5$ is moderate; $5\le\left|\ln \mathcal{B}_{ij}\right|<10$ is strong; and if $\left|\ln \mathcal{B}_{ij}\right|\ge 10$, the evidence is decisive. It is worth noting that a positive value of $\ln \mathcal{B}_{ij}$ represents a preference for model $i$ over model $j$.

In figure~\ref{fig6}, we show the Bayes factors $\ln \mathcal{B}_{ij}$ for the PEDE and PEDE+$w_{\rm dm}$ models relative to the $\Lambda$CDM model, based on the current observational data. Here, $i$ denotes the PEDE or PEDE+$w_{\rm dm}$ model and $j$ denotes the $\Lambda$CDM model. It is worth emphasizing that negative values indicate a preference for the $\Lambda$CDM model. The CMB, CMB+DESI, CMB+DESI+DESY5, and CMB+DESI+PantheonPlus data yield Bayes factors of $\ln \mathcal{B}_{ij}$ = 4.29, $-$4.78, $-$13.35 , and $-$10.68 for PEDE, and $\ln \mathcal{B}_{ij}$ = 3.08, $-$9.17, $-$17.93, and $-$14.21 for PEDE+$w_{\rm dm}$, relative to $\Lambda$CDM, respectively. We find that, when using CMB data alone, there is a moderate preference for both the PEDE and PEDE+$w_{\rm dm}$ models relative to the $\Lambda$CDM model. When the DESI, DESY5, and PantheonPlus data are included, the current data combination strongly prefers $\Lambda$CDM over the PEDE and PEDE+$w_{\rm dm}$ models. In addition, in this case, the PEDE+$w_{\rm dm}$ model is more strongly disfavored than the PEDE model.

\section{Conclusion}\label{sec4}

Dark matter dominates the matter content of the universe, however, aside from its density, our understanding of its properties remains limited. In most cosmological models, dark matter is assumed to be pressureless or cold, meaning that the dark matter EoS parameter is zero. Although this hypothesis is based on the abundance of CDM in the universe, there is no fundamental theoretical reason to assume that the dark matter EoS parameter is exactly zero. Recently, Kumar \emph{et al.} \cite{Kumar:2025etf} investigated the $\Lambda$CDM framework by treating the dark matter EoS parameter as a free parameter, and found a deviation from zero at the $2\sigma$ level. In this work, we explore whether a non-zero dark matter EoS parameter is favored by the current cosmological observations within the PEDE model, due to its simplicity and ability to alleviate the Hubble tension. Furthermore, we investigate the impact of a non-zero dark matter EoS parameter on the Hubble tension.

We discuss the constraints on the PEDE+$w_{\rm dm}$ model using CMB alone, and then gradually investigate the effects of other cosmological probes as they are added to CMB. We find that using CMB data alone cannot provide strong constraints on the PEDE+$w_{\rm dm}$ model due to the degeneracy among parameters, and no evidence for a deviation of $w_{\rm dm}$ from zero is found. When DESI data are added to CMB, we observe that $w_{\rm dm}$ deviates from zero at slightly more than the $1\sigma$ level, with $w_{\rm dm} = -0.00036\pm0.00034$. Furthermore, the inclusion of low-redshift data significantly improves the precision of parameter constraints, as it effectively breaks the parameter degeneracy. When combining the DESY5 and DESI data with CMB, the constraint on $w_{\mathrm{dm}}$ is $-0.00093 \pm 0.00032$, showing a larger deviation of $w_{\rm dm}$ from zero at approximately the $3\sigma$ level. However, once DESY5 is replaced with PantheonPlus, the deviation of $w_{\rm dm}$ from zero decreases from approximately $3\sigma$ to $2\sigma$. Further analysis suggests that the shift in $w_{\rm dm}$ from CMB+DESI+DESY5 could potentially be due to the presence of unidentified supernova systematics or the inconsistencies in $\Omega_{\mathrm{m}}$ between CMB and DESY5. 

It is worth noting that due to the strong positive correlation between $w_{\mathrm{dm}}$ and $H_0$, a more negative value of $w_{\mathrm{dm}}$ leads to a lower value of $H_0$. For example, the CMB+DESI+DESY5 data yields $H_0 = 71.34\pm 0.31~\rm km~s^{-1}~Mpc^{-1}$ in the PEDE model and $H_0 = 70.62\pm 0.44~\rm km~s^{-1}~Mpc^{-1}$ in the PEDE+$w_{\rm dm}$ model. As a result, the Hubble tension increases from $1.57\sigma$ in the PEDE model to $2.14\sigma$ in the PEDE+$w_{\rm dm}$ model. The results obtained here indicate that we cannot simultaneously obtain evidence for non-CDM and alleviate the Hubble tension within the framework of this study.   
 
We fit the CMB power spectrum and observe that the amplitude at lower multipoles is slightly suppressed in the PEDE+$w_{\rm dm}$ model relative to the $\Lambda$CDM model, resulting in a better agreement with the CMB data. Additionally, the best-fit predictions of the PEDE+$w_{\rm dm}$ model greater success than the $\Lambda$CDM model in explaining $D_{\mathrm{M}}(z)/(r_{\mathrm{d}}\sqrt{z})$ at $z = 0.71$. Finally, in order to facilitate a quantitative comparison for various models, we calculate the Bayesian evidence for the PEDE and PEDE+$w_{\rm dm}$ models utilizing current observational data. When using CMB data alone, there is a moderate preference for both the PEDE and PEDE+$w_{\rm dm}$ models relative to the $\Lambda$CDM model. When DESI, DESY5, and PantheonPlus data are included, the current data combination strongly prefers $\Lambda$CDM over the PEDE and PEDE+$w_{\rm dm}$ models. Our results indicate that considering a non-CDM component within the PEDE model is not feasible. However, incorporating a free dark matter EoS parameter presents an interesting and worthwhile scenario for exploration within other frameworks.

\section*{Acknowledgments}
We thank Jing-Zhao Qi, Jia-Le Ling, and Sheng-Han Zhou for their helpful discussions. This work was supported by the National SKA Program of China (Grants Nos. 2022SKA0110200 and 2022SKA0110203), the National Natural Science Foundation of China (Grants Nos. 12533001, 12575049, and 12473001), the China Manned Space Program (Grant No. CMS-CSST-2025-A02), the National 111 Project (Grant No. B16009), and Guangdong Basic and Applied Basic Research Foundation (Grant No. 2024A1515012573).

\bibliography{main}
\bibliographystyle{JHEP}
\end{document}